\documentclass[journal,transmag]{IEEEtran}
\usepackage{hyperref,color,multirow}
\usepackage[top=0.7in, left=0.65in]{geometry}
\usepackage{graphicx}
\usepackage[font=footnotesize]{caption}

\usepackage{subcaption} 

\usepackage{pgfplots}
\usepackage{tikz}
\usetikzlibrary{patterns}
\usetikzlibrary{spy}
\usepackage{listings}
\usepackage{pgfkeys}
\usetikzlibrary{intersections}
\usepackage{amssymb}
\usepackage{comment}

\usepackage{amsmath}

\definecolor{mc}{rgb}{0.19,0.25,0.45} 
\definecolor{mc2}{rgb}{.54, .65, .80} 
\definecolor{myblue}{rgb}{.2,0.45,0.5} 
\definecolor{hcolor}{HTML}{FF6000}
\definecolor{acolor}{HTML}{2C83C5}
\definecolor{alert}{HTML}{E73811}

\definecolor{myblue}{rgb}{.2,0.45,0.5} 
\definecolor{myorange}{rgb}{0.78,0.6,0.3}
\definecolor{mygreen}{rgb}{.2,0.38,0.16}
\definecolor{myalert}{rgb}{0.97,0.09,0.21}

\renewcommand{\vec}[1]{\boldsymbol{#1}} 

\newcommand{\paren}[1]{\left(#1\right)}
\newcommand{\parenangle}[1]{\left\langle#1\right\rangle}

\newcommand{\volInt}[3]{\paren{#1 \, , #2}_{#3}}
\newcommand{\volIntBig}[3]{\big(#1 \, , #2\big)_{#3}}

\newcommand{\surInt}[3]{\parenangle{#1 \, , #2}_{#3}}

\makeatletter
\newcommand\Label[1]{&\refstepcounter{equation}\text{(\theequation)}\ltx@label{#1}&}
\makeatother

\newlength\Colsep
\newcommand{\curl}{\text{\textbf{curl}}\, }
\newcommand{\curlOnly}{\text{\textbf{curl}}}

\renewcommand{\b}{\vec b}
\renewcommand{\a}{\vec a}

\newcommand{\h}{\vec h}

\newcommand{\e}{\vec e}

\renewcommand{\j}{\vec j}

\newcommand{\dt}{\partial_t}
\newcommand{\ec}{e_{\text{c}}}
\newcommand{\jc}{j_{\text{c}}}

\renewcommand{\O}{\Omega}
\newcommand{\Oa}{\Omega_a}
\newcommand{\Oh}{\Omega_h}
\newcommand{\Ohc}{\Omega_{h,\text{c}}}
\newcommand{\Ohcc}{\Omega_{h,\text{c}}^{\text{C}}}

\newcommand{\Oc}{\Omega_{\text{c}}}
\newcommand{\Occ}{\Omega_{\text{c}}^{\text{C}}}
\newcommand{\Om}{\Omega_{\text{m}}}
\newcommand{\Omc}{\Omega_{\text{m}}^{\text{C}}}
\newcommand{\Os}{\Omega_{\text{s}}}
\newcommand{\Gm}{\Gamma_\text{m}}
\newcommand{\af}{$a$-formulation\ }

\newcommand{\hf}{$h$-formulation\ }
\newcommand{\hpf}{$h$-$\phi$-formulation\ }

\newcommand{\ajf}{$a$-$j$-formulation\ }
\newcommand{\hpaf}{$h$-$\phi$-$a$-formulation\ }

\newcommand{\afOnly}{$a$-formulation}
\newcommand{\abfOnly}{$\bar a$-formulation}

\newcommand{\hfOnly}{$h$-formulation}
\newcommand{\hpfOnly}{$h$-$\phi$-formulation}
\newcommand{\hafOnly}{$h$-$a$-formulation}
\newcommand{\hbfOnly}{$h$-$\phi$-$b$-formulation}
\newcommand{\ajfOnly}{$a$-$j$-formulation}
\newcommand{\hpafOnly}{$h$-$\phi$-$a$-formulation}

\newcommand{\afc}{$a$\ }
\newcommand{\abfc}{$\bar a$\ }

\newcommand{\hfc}{$h$\ }
\newcommand{\hpfc}{$h$-$\phi$\ }
\newcommand{\hafc}{$h$-$a$\ }
\newcommand{\hbfc}{$h$-$\phi$-$b$\ }
\newcommand{\ajfc}{$a$-$j$\ }
\newcommand{\hpafc}{$h$-$\phi$-$a$\ }

\newcommand{\hsp}{\mathcal{H}}
\newcommand{\asp}{\mathcal{A}}

\usepackage{eso-pic}

\begin{document}
\title{What Formulation Should One Choose for Modeling a 3D HTS Motor Pole with Ferromagnetic Materials?}
\author{\IEEEauthorblockN{Julien Dular\IEEEauthorrefmark{1}, Kévin Berger\IEEEauthorrefmark{2},
Christophe Geuzaine\IEEEauthorrefmark{1}, and
Benoît Vanderheyden\IEEEauthorrefmark{1}}
\IEEEauthorblockA{\IEEEauthorrefmark{1}University of Liège, Institut Montefiore, Liège, Belgium, Julien.Dular@uliege.be}{\IEEEauthorrefmark{2}Université de Lorraine, GREEN, Nancy, France}
\thanks{}}

\IEEEtitleabstractindextext{%
\begin{abstract}
We discuss the relevance of several finite-element formulations for nonlinear systems containing high-temperature superconductors (HTS) and ferromagnetic materials (FM), in the context of a 3D motor pole model. The formulations are evaluated in terms of their numerical robustness and efficiency. We propose a coupled \hpaf as an optimal choice, modeling the problem with an \af in the FM and an \hpf in the remaining domains. While maintaining a low number of degrees of freedom, the \hpaf guarantees a robust resolution and strongly reduces the number of iterations required for handling the nonlinearities of HTS and FM compared to standard formulations.
\end{abstract}

\begin{IEEEkeywords}
Finite element analysis, high-temperature superconductors, magnetic materials, coupled formulations.
\end{IEEEkeywords}}

\maketitle

\AddToShipoutPicture*{
    \footnotesize\sffamily\raisebox{0.8cm}{\hspace{1.5cm}\fbox{
        \parbox{\textwidth}{
            \copyright~2022
                IEEE. Personal use of this material is permitted. Permission from IEEE
                must be obtained for all other uses, in any current or future media, 
                including reprinting/republishing this material for advertising or
                promotional purposes, creating new collective works, for resale or
                redistribution to servers or lists, or reuse of any copyrighted
                component of this work in other works.
            }
        }
    }
}

\thispagestyle{empty}
\pagestyle{empty} 


\section{Introduction}

\IEEEPARstart{M}{odeling} the magnetic response of high-temperature superconductors (HTS) is important for many applications. One of the main modeling tools is the finite element method with the $E$-$J$ power law in HTS \cite{grilli2013development}. The resulting system of equations is strongly nonlinear and requires one to choose the formulation carefully to obtain both accurate results and fast resolutions. Coupling HTS with ferromagnetic materials (FM) introduces additional difficulties. In particular, the power law in HTS and magnetic law in FM are most efficiently solved with distinct formulations \cite{Dular2019}.

In this work, we compare the computational cost of a number of formulations with both HTS and FM for a 3D problem.

In section \ref{secProblem}, we describe the 3D problem and define the nonlinear constitutive laws. In section~\ref{secFormulations}, we introduce the different formulations that will be compared. We first present classical $h$- and $b$-conform formulations. Then, we propose four different coupled formulations that involve the material laws in an efficient manner. To the best of our knowledge, the \ajf (in 3D), \hbfOnly, and \hpaf are original contributions. In section~\ref{secResults}, we compare the formulations in terms of the associated computational time.

\section{Magnet Pole Problem}\label{secProblem}

\begin{figure}[h!]
\begin{center}
\includegraphics[width=0.55\linewidth]{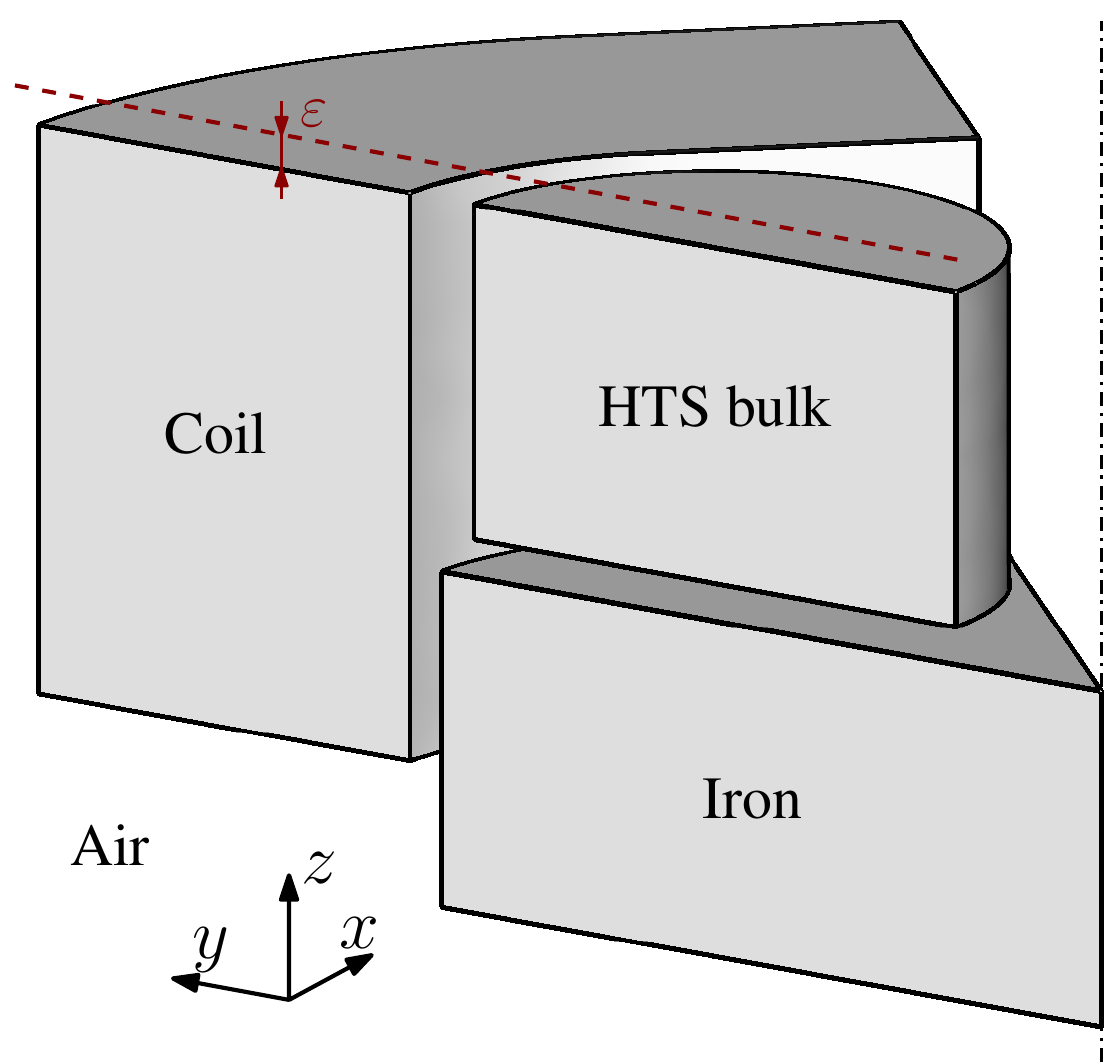}
\caption{Geometry of one eighth of the 3D geometry. The red dashed line is where the magnetic flux density is represented in Fig.~\ref{b_onLine}, at $\varepsilon = 2$~mm above the HTS bulk and coil top surfaces, in the symmetry plane. Bulk height: 17.7~mm. Bulk radius: 15~mm. Numerical domain radius: 270~mm.}
\label{geometry}
\end{center}
\end{figure}

We consider four HTS bulks placed on top of an iron substrate, and magnetized by an inducting coil. Exploiting symmetry, one eighth of the geometry is modeled, as shown in Fig.~\ref{geometry}. This geometry is relevant in applications, e.g., \cite{berger20183}.

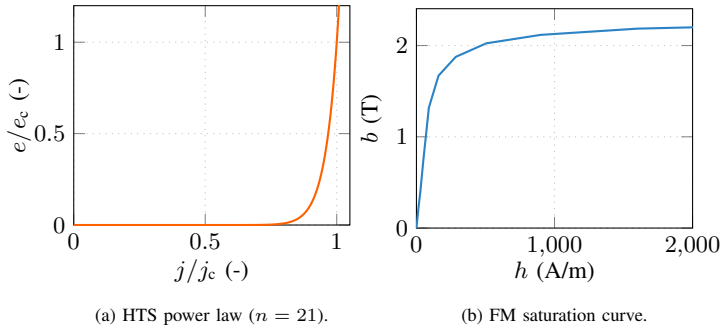
\begin{figure}[h!]
        \centering
	\begin{subfigure}[b]{0.49\linewidth}  
     \centering 
	\begin{tikzpicture}[trim axis left, trim axis right][font=\small]
\begin{axis}[
    width=1.2\linewidth,
    height=4.5cm,
    grid = both,
    grid style = dotted,
    xmin=0, 
    xmax=1.05,
    ymin=-0.0, 
    ymax=1.2,
	xtick scale label code/.code={},
	ytick scale label code/.code={},
	xlabel={$j/j_\text{c}$ (-)},
    ylabel={$e/e_\text{c}$ (-)},
    ylabel style={yshift=-1.6em},
    xlabel style={yshift=0.6em},
    legend style={at={(0.03, 0.99)}, anchor=north west, draw=none}
    ]
        \addplot[hcolor, thick, domain=0:1.05, samples=201]
		{x^21};
    \end{axis}
	\end{tikzpicture}%
    \caption{HTS power law ($n=21$).}
    \label{nonlinearities_HTS}
    \end{subfigure}
        \hfill
        \begin{subfigure}[b]{0.49\linewidth}
            \centering
	\begin{tikzpicture}[trim axis left, trim axis right][font=\small]
  	\begin{axis}[
	tick label style={/pgf/number format/fixed},
    width=1.2\linewidth,
    height=4.5cm,
    grid = both,
    grid style = dotted,
    xmin=-0, 
    xmax=2000,
    ymin=0, 
	xtick scale label code/.code={},
    xtick={0, 1000, 2000},
	ytick scale label code/.code={},
	ylabel={$b$ (T)},
    xlabel={$h$ (A/m)},
    ylabel style={yshift=-2em},
    xlabel style={yshift=0.6em},
    legend style={at={(0.03, 0.99)}, anchor=north west, draw=none}
    ]
        \addplot[acolor, thick] 
    table[x=h,y=b]{b_of_h.txt};
    \end{axis}
	\end{tikzpicture}%
            \caption{FM saturation curve.}
            \label{nonlinearities_FM}
        \end{subfigure}
        \caption{Nonlinear laws involved in the problem.}
        \label{nonlinearities}
\end{figure}

We model the magnetic response of the system with the magneto-quasistatic approximation, i.e., we neglect the displacement current in Maxwell's equations. In HTS, $\mu = \mu_0$ and we assume a power law for the electrical resistivity \cite{douine2015determination}:
\begin{equation}\label{eqn_contitutiveje}
\rho = \frac{\ec}{\jc}\paren{ \frac{\|\j\|}{\jc}}^{n-1},
\end{equation}
with $\ec = 10^{-4}$ V/m, $\jc$ the critical current density and $n$ the power index.
Both $\jc$ and $n$ depend on the norm of the local magnetic flux density \cite{douine2015determination}:
\begin{align}
\jc(\b) = \frac{j_\text{c0}}{1+ \|\b\|/b_0}, \quad n(\b) = n_1 + \frac{n_0-n_1}{1+ \|\b\|/b_0},
\end{align}
with $j_\text{c0} = 5\times 10^8$ A/m$^2$, $b_0 = 0.5$ T, $n_0 = 21$, and $n_1 = 5$ (representative values for YBCO pellets). The iron is assumed non-conducting and its permeability $\mu$ follows a saturation law based on experimental data (saturation at $\approx 2.2$ T). The nonlinearities associated with both materials are depicted in Fig.~\ref{nonlinearities}.

The inducting coil contains $55.5$ turns and carries an imposed current $I_\text{s}$ defined by 
\begin{align}
I_\text{s}(t) = \left\{\begin{aligned}
&I_\text{max}\ \sin(\pi t /2T),\ &t<T,\\
&I_\text{max}\ e^{-(t-T)/\tau},\ &t>T,\end{aligned}\right.
\end{align}
with $I_\text{max} = 2$ kA, $T=2$ ms, and $\tau = 10$ ms. Eddy currents in the coil are neglected. We model the response of the system from a virgin state at $t=0$ (zero-field cooled) to $t=10\,T$.

\section{Finite Element Formulations}\label{secFormulations}

We now present the formulations that will be compared. First, we introduce standard $h$-conform and $b$-conform formulations. Motivated by observations on their numerical behavior, we then introduce coupled formulations in order to improve their efficiency. A summary of the considered formulations is proposed in Table~\ref{tableFormulations}.

The modeled domain $\O$ is decomposed into a conducting part, $\Oc$, containing the HTS bulk, and a non-conducting part $\Occ$, containing the air, the iron, and the coil. In addition, the iron is referred to as $\Om$, and its complementary $\Omc$. The coil is denoted by $\Os$.

The geometry is discretized with a finite element mesh. In the following, unknown and test functions are directly defined in discrete function spaces associated with the mesh.

\begin{table*}[t]
\caption{Description of the different formulations}
\vspace{-0.35cm}
\begin{center}
\begin{tabular}{l c l l c}
\hline
 & NL laws & Function space & Number of DOFs & $\sigma\neq 0$ in $\Occ$?\\
\hline
\hfc & $\rho$, $\mu$ & $\h \in \hsp(\O) = \{\h \in H(\O)\}$ & Edges in $\O$ & Yes\\
\hpfc & $\rho$, $\mu$ & $\h \in \hsp_\phi(\O) = \{\h \in H(\O)\ |\ \curl \h = \vec 0 \text{ in } \Occ\}$ &  Edges in $\Oc$ + Nodes in $\Occ$ & No\\
\abfc & $\sigma$, $\nu$ & $\a \in \bar \asp(\O) = \{\a\in H(\O)\}$ & Edges in $\O$ & (Yes)$^*$\\
\afc & $\sigma$, $\nu$ & $\a \in \asp(\O) = \{\a\in H(\O)\ |\ \text{co-tree gauge in }\Occ\}$ & Edges in $\Oc$ + Facets in $\Occ$ & No\\
\hafc & $\rho$, $\nu$ & $\h \in \hsp_\phi(\Oc)$, $\a \in \asp(\Occ)$& Edges in $\Oc$ + Facets$^{\dagger}$ in $\Occ$ & No\\
\hpafc & $\rho$, $\nu$ & $\h \in \hsp_\phi(\Omc)$, $\a \in \asp(\Om)$&Edges in $\Ohc$ + Nodes$^{\dagger}$ in $\Ohcc$ + Facets in $\Om$ & No\\
\hbfc & $\rho$, $\nu$ & $\h \in \hsp_\phi(\O)$, $\b \in (H_3(\Om))^3$& Edges in $\Oc$ + Nodes in $\Occ$ + Volumes ($\times3$) in $\Om$ & No\\
\ajfc & $\rho$, $\nu$ & $\a \in \asp(\O)$, $\j \in \asp(\Oc)$ & Edges ($\times 2$) in $\Oc$ + Facets in $\Occ$ & No\\
\hline
\end{tabular}
\end{center}
NL stands for nonlinear. $H(\O)$ is the finite dimensional subset of the curl-conform space $H(\curlOnly; \Omega)$ generated by Whitney edge elements on the mesh in $\O$ \cite{bossavit1988whitney}. $(H_3(\Om))^3$ is the space of piecewise constant vector functions (3 components) per element in $\Om$. $^*$For the \abfOnly, choosing $\sigma =0$ in $\Occ$ makes the system singular, but this is not necessarily an issue as some linear solvers do not require uniqueness of the solution. $^{\dagger}$For surface-coupled formulations, a local enrichment is necessary on $\Gm$ to guarantee stability \cite{dularStability}. In 3D, this adds twice the number of facets on $\Gm$ as DOFs (to enrich $\asp$, our choice for the \hafOnly), or once the number of edges on $\Gm$ (to enrich $\hsp_\phi$, our choice for the \hpafOnly).
\label{tableFormulations}
\end{table*}

\subsection{Standard $h$-conform Formulations}

We consider common $h$-conform formulations that are written as a weak form of Faraday's law. The unknown field is the magnetic field $\h$, sought in a specific function space, such that
\begin{align}
\volIntBig{\dt (\mu \h)}{\h'}{\O} + \volIntBig{\rho\,\curl \h}{\curl \h'}{\Oc}  = 0
\end{align}
holds for all $\h'$ in the same space (for conciseness, we consider homogeneous boundary conditions). The choice of the function space determines the resulting number of degrees of freedom (DOFs). We present here two options.

First, in the full \hfOnly, the magnetic field is discretized with edge functions on the whole domain, and a spurious resistivity $\rho_\text{s}$ is introduced in $\Occ$ \cite{grilli2013development}. Despite leading to more unknowns and ill-conditioned matrices \cite{dlotko2019fake}, this approach is still popular in proprietary softwares, e.g., COMSOL.

A second option is to strongly impose a zero current density in non-conducting regions by carefully defining the function space, leading to the well-known \hpf \cite{dular2000dual}. Multiply connected subdomains, as is the case here for the complementary of the coil, are handled with discontinuous or cohomology basis functions. Here, a precomputed source field $\h_{\text{s}}$ reproducing the source current density in the coil $\Os$ is included in the space of $\h$.


Note that other $h$-conform formulations also exist, such as the $t$-$\omega$-formulation \cite{nakata1988comparison}. Because it involves the nonlinearities in the same manner ($\rho$ and $\mu$), the overall conclusions are expected to be similar to those for $h$- or $h$-$\phi$-formulations.

\subsection{Standard $b$-conform Formulations}

Standard $b$-conform formulations are expressed as a weak form of Ampère's law. The unknown field is a vector potential $\a$, with $\curl \a = \b$, sought in a given function space, such that
\begin{align}
\volInt{\nu\, \curl \a}{\curl \a'}{\O} + \volInt{\sigma\,\dt \a}{\a'}{\Oc} = \volInt{\j_\text{s}}{\a'}{\Os}
\end{align}
holds for all $\a'$ in the same space, where $\j_\text{s}$ is the imposed current density in the coil, $\nu = 1/\mu$ is the reluctivity, and $\sigma = 1/\rho$ is the conductivity. In our problem, $\e = -\dt \a$ \cite{Dular2019}.

The vector potential is not unique in $\Occ$, and can be gauged. For example, a co-tree gauge reduces the number of DOFs, and leads to what we call the \af \cite{dular2000dual}. As with the $h$-conform formulations, we can also introduce a spurious non-zero conductivity $\sigma_\text{s}$ in $\Occ$, and hence avoid further gauging steps. We refer to this choice as the \abfOnly. Note that gauging is not mandatory with $\sigma_\text{s} = 0$. The ungauged system is singular, but depending on the linear solver, this is not necessarily an issue, e.g., for some iterative solvers.

\subsection{Surface-Coupled Formulations}

The important differences between the $h$-conform and $b$-conform formulations are the involved nonlinear laws. It has been observed that $\rho$ and $\nu$ are easier to handle than $\sigma$ and $\mu$, mostly due to the shape of the associated constitutive relationship \cite{Dular2019}. Therefore, because neither one of the above formulations involves both $\rho$ and $\nu$, there is a motivation for investigating coupled formulations.

In this section, we present formulations with surface coupling. The domain $\O$ is decomposed in two parts: $\Oh$, to be solved with the \hpfOnly, and $\Oa$, to be solved with the \afOnly. The HTS is always put in $\Oh$ so as to involve $\rho$ and the FM is always put in $\Oa$ so as to involve $\nu$. The remaining domains can be either put in $\Oh$ or in $\Oa$. We consider two choices. In the \hpafOnly, only the FM domain $\Om$ belongs to $\Oa$, and an \hpf is used in $\Omc$. In the \hafOnly, we place all non-conducting domains $\Occ$ in $\Oa$ \cite{Dular2019, bortot2020coupled}.

In both cases, the coupling is performed via the common boundary $\Gm$ and the formulation amounts to finding $\h$ and $\a$ such that
\begin{align}
\begin{aligned}&\volIntBig{\mu_0\dt \h}{\h'}{\Oh} + \volIntBig{\rho\,\curl \h}{\curl \h'}{\Ohc}= \surInt{\dt \a \times\vec n_{\Oa}}{\h'}{\Gamma_\text{m}}\\
&\volInt{\nu\, \curl \a}{\curl \a'}{\Oa} - \volInt{\j_\text{s}}{\a'}{\O_\text{s}\cap \Oa}= \surInt{\h\times\vec n_{\Oa}}{\a'}{\Gamma_\text{m}}
\end{aligned}
\end{align}
hold for all $\h'$ and $\a'$, with $\Ohc$ the conducting part of $\Oh$ (here the HTS domain), and $\vec n_{\Oa}$ the outer normal of $\Oa$. For this mixed formulation to be stable, either $\h$ or $\a$ should be enriched with second-order elements on $\Gm$ if linear elements are used elsewhere \cite{dularStability}.

\subsection{Volume-Coupled Formulation}

A second kind of mixed formulations is obtained when auxiliary fields are added in the volume of a region. For example, the \hpf is not optimal because of the shape of the permeability law $\mu$, which is difficult to handle in a robust manner. To introduce the reluctivity $\nu$ instead, the formulation can be modified as follows: find $\h$ and $\b$ such that
\begin{align}
&\volIntBig{\dt \b}{\h'}{\Om} + \volIntBig{\mu_0\dt \h}{\h'}{\Omc}+ \volIntBig{\rho\,\curl \h}{\curl \h'}{\Oc}  = 0\notag\\
&\volInt{\nu\b}{\b'}{\Om} - \volInt{\h}{\b'}{\Om} = 0
\end{align}
holds for all $\h'$ and $\b'$, where the auxiliary flux density $\b$ is only defined in $\Om$ to minimize the additional DOFs. The function space for $\b$ is chosen to be piecewise constant per volume element. 
We refer to this formulation as the \hbfOnly.

The analogous approach starting from the \af consists in introducing $\j$ as an auxiliary field and solve the following mixed problem, so as to involve the resistivity $\rho$: find $\a$ and $\j$ such that
\begin{align}
\begin{aligned}
&\volInt{\nu\, \curl \a}{\curl \a'}{\O} - \volInt{\j}{\a'}{\Oc} = \volInt{\j_\text{s}}{\a'}{\Os}\\
&\volInt{\rho \j}{\j'}{\Oc} + \volInt{\dt \a}{\j'}{\Oc} = 0
\end{aligned}
\end{align}
holds for all $\a'$ and $\j'$, with $\j$ defined in the same space than $\a$ in $\Oc$. This formulation is close to the $a$-$v$-$j$-formulation proposed in \cite{Stenvall2010a} for 2D problems. Here, because the problem does not involve a scalar electric potential, we refer to this formulation as the \ajfOnly.

\section{Comparison on the 3D Problem}\label{secResults}

We implemented the eight formulations in GetDP \cite{getdp}. Model files are available at \url{www.life-hts.uliege.be}. Models were run on one 2.9 GHz AMD Epyc Rome 7542 CPU.

\subsection{Implementation Details}

Formulations are integrated over time with $128$ time steps. Linear systems are solved with the direct sparse solver MUMPS. 
An iterative Newton-Raphson method is used for $\rho$ and $\nu$. For $\sigma$ and $\mu$, we were only able to obtain a robust method by using Picard fixed point iterations. Even though the use of relaxation factors on the Newton-Rapshon iterations sometimes help, we have not found a robust set of numerical parameters allowing convergence in all cases. Iterations typically enter cycles that are difficult to avoid \cite{Dular2019}.

The convergence criterion is based on the instantaneous power $P=\volInt{\dt \b}{\h}{\O}+\volInt{\j}{\e}{\Oc}$. Iterations stop when the relative change of $P$ is smaller than $10^{-6}$ (or $10^{-5}$ in case of fixed point iterations) in each of the sub-domains.

For the \hfOnly, the spurious resistivity in air is fixed to $\rho_\text{s} =10^{-3}$ $\O$m. For the \abfOnly, $\sigma_\text{s} = 1$ S/m. With these values, we have not observed any significant impact of the spurious parameters on the numerical solution quality.

\subsection{Results}

\begin{figure}[h!]
\centering
        \begin{subfigure}[b]{0.99\linewidth}  
\centering
	\begin{tikzpicture}[trim axis left, trim axis right][font=\small]
 	\begin{axis}[
 	tick scale binop=\times,
    width=1\linewidth,
    height=3.5cm,
    grid = both,
    grid style = dotted,
    xmin=-5, 
    xmax=120,
    ymin=0, 
    ymax=0.7,
	ylabel={$\|b\|$ (T) },
    ylabel style={yshift=-1.5em},
    legend style={at={(0.99, 0.99)}, anchor=north east, draw=none}
    ]
    \addplot[hcolor, thick] 
    table[x=position_y,y=bnorm_h_t1]{bLine2.txt};
        \addplot[acolor, thick] 
    table[x=position_y,y=bnorm_a_t1]{bLine2.txt};
    \legend{\hpf, \af}
    \end{axis}
	\end{tikzpicture}%
	\end{subfigure}
        \hfill
        \begin{subfigure}[b]{0.99\linewidth}  
\centering
		\begin{tikzpicture}[trim axis left, trim axis right][font=\small]
 	\begin{axis}[
 	tick scale binop=\times,
    width=1\linewidth,
    height=3.7cm,
    grid = both,
    grid style = dotted,
    xmin=-5, 
    xmax=120,
    ymin=-0.24, 
    ymax=1.3,
	ylabel={$\|b\|$ (T) },
    xlabel={Position (mm)},
    ylabel style={yshift=-1.5em},
    xlabel style={yshift=0.7em},
    legend style={at={(0.99, 0.99)}, anchor=north east, draw=none}
    ]
    \addplot[hcolor, thick] 
    table[x=position_y,y=bnorm_h_t2]{bLine2.txt};
        \addplot[acolor, thick] 
    table[x=position_y,y=bnorm_a_t2]{bLine2.txt};
         \addplot[gray] 
    coordinates {(0,-1) (0, 0)};
     \addplot[gray] 
    coordinates {(0,0) (30, 0)}; 
     \addplot[gray] 
    coordinates {(30,0) (30, -1)}; 
     \addplot[gray] 
    coordinates {(34,0) (34, -1)}; 
        \addplot[gray] 
    coordinates {(34,0) (57.11, 0)}; 
         \addplot[gray] 
    coordinates {(57.11,0) (57.11, -1)}; 
    \node[] at (axis cs: 15,-0.12) {\textcolor{gray}{HTS}};
    \node[] at (axis cs: 45.555,-0.12) {\textcolor{gray}{Coil}};
    \legend{\hpf, \af}
    \end{axis}
	\end{tikzpicture}%
		\end{subfigure}
\caption{Norm of the magnetic flux density at $\varepsilon = 2$~mm above the system, along the dashed line represented in Fig.~\ref{geometry}. The upper plot is at $t=0.25\,T$ and the lower plot at $t=3.5\,T$. Curves from $h$-, $h$-$\phi$-$a$-, $h$-$\phi$-$b$-formulations are indistinguishable from that of the \hf ($h$-conform field in air). The same is true for curves from $\bar a$-, $h$-$a$-, $a$-$j$-formulations in comparison with those from the \af ($b$-conform field in air).}
\label{b_onLine}
\end{figure}
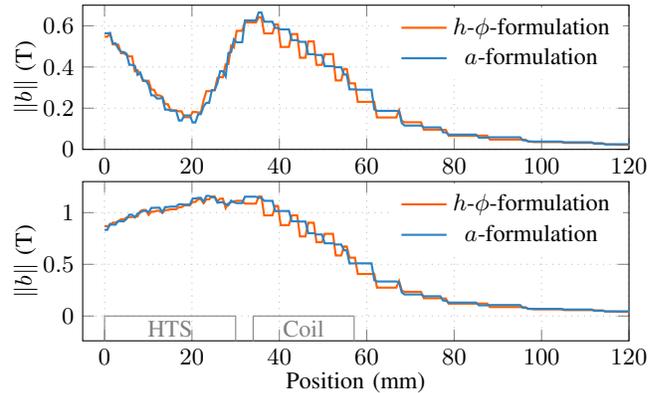

\begin{figure}[h!]
\begin{center}
\includegraphics[width=0.9\linewidth]{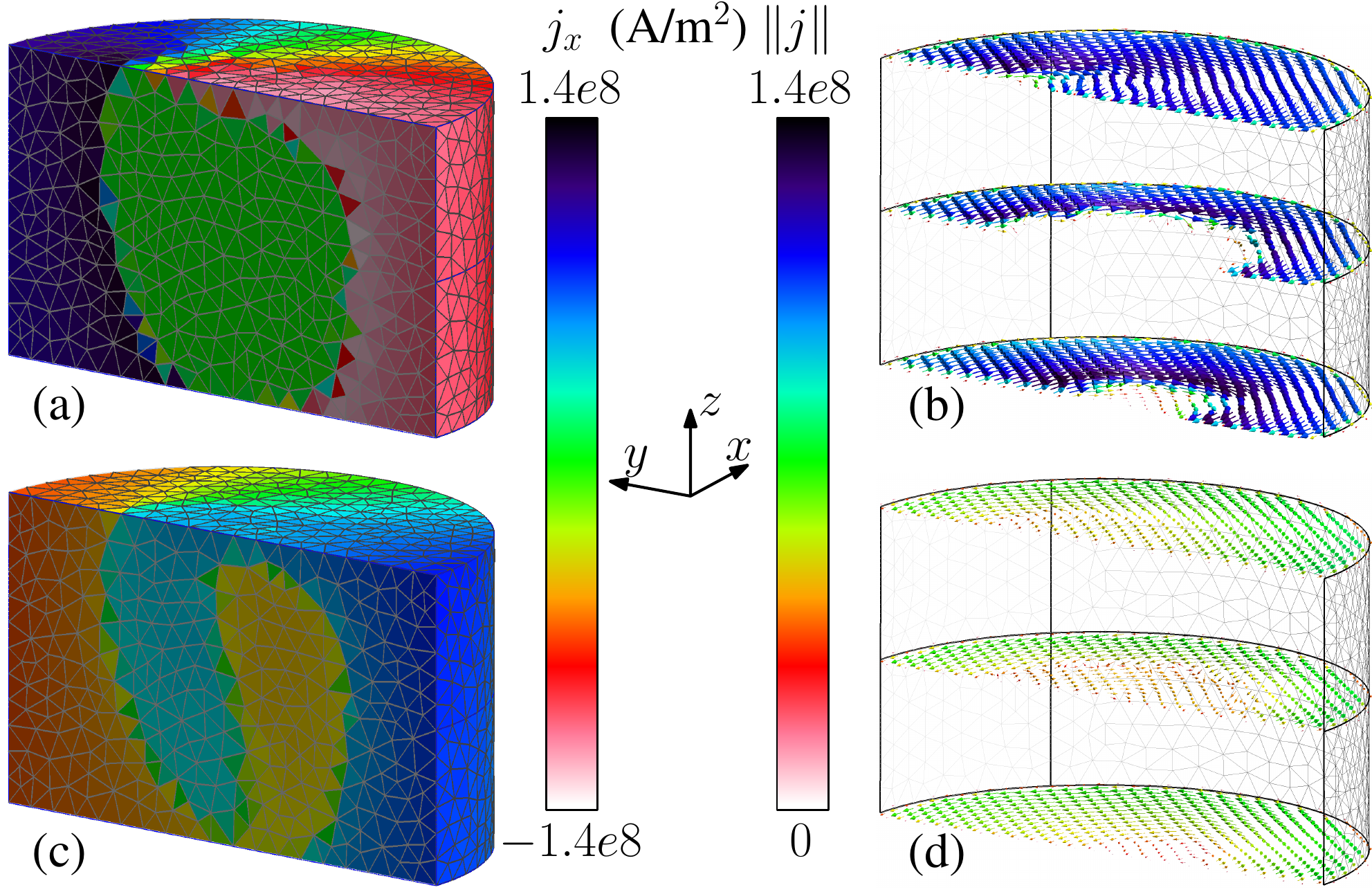}
\caption{Current density  from the \hpaf in the bulk during the magnetizing pulse:  (a)-(b) at $t=0.25\,T$, and (c)-(d) during relaxation at $t=3.5\,T$. (a)-(c) $x$-component $j_x$ in the bulk, and (b)-(d) full vector $\j$ in three planes.}
\label{b_onPlane}
\end{center}
\end{figure}

We simulate the problem with the eight formulations on the same mesh. Global and local solutions agree with each other. The total hysteresis loss in the bulk is given in Table~\ref{tableResults}, the difference between the values is at most 1\%. The norm of $\b$ along the dashed line of Fig.~\ref{geometry} is represented in Fig.~\ref{b_onLine} for the $a$- and $h$-$\phi$-formulations. All other formulations yield results that are visually indistinguishable from these two formulations. Results also match inside the HTS and FM. The current density in the HTS is represented in Fig.~\ref{b_onPlane}.

A good accuracy can be achieved with all formulations. However, the computational cost associated with each of them is not equivalent, see Table~\ref{tableResults}. First, the number of DOFs is strongly affected by the choice of function spaces. When possible, is it always preferable to introduce a magnetic scalar potential $\phi$ and to gauge the magnetic vector potential $\a$ in $\Occ$. Also, using $\phi$ in air instead of $\a$ in surface-coupled formulations leads to fewer DOFs.

\begin{table}[!h]
\caption{Comparison of the different formulations}
\vspace{-0.25cm}
\begin{center}
\begin{tabular}{l c c c c c}
\hline
 & HTS loss (J) & \# DOFs & \# iterations & Time/it. & Total time \\
\hline
\hfc & 6.35 & 35,532 & 4,057 & 3.3s & 3h42\\
\hpfc & 6.36 & 12,172 & 3,937 & 1.4s & 1h33\\
\abfc & 6.38 & 29,010 & 2,955 & 3.1s & 2h33\\ 
\afc & 6.39 & 26,964 & 3,147 & 2.1s & 1h48\\
\hafc & 6.31 & 32,045 & 1,124 & 2.7s & 0h50\\
\hpafc & 6.33 & 15,776 & 1,108 & 2.1s & 0h39\\
\hbfc & 6.37 & 20,821 & 1,104 & 3.2s & 0h58\\
\ajfc & 6.34 & 36,019 & 2,225 & 3.6s & 2h15\\
\hline
\end{tabular}
\end{center}
Performance figures for the 128 time steps of the eight formulations with linear elements (except on the coupling boundary for coupled formulations where second order elements are used). HTS loss is the total hysteresis loss in the HTS bulk from $t=0$ to $t=10\,T$. Results differ by maximum $1$\%. Picard fixed point iterations were used for the first four formulations, which explains the large associated number of iterations. With more efficient method such as Newton-Raphson iterations (with or without relaxation factors), we have not obtained robust behaviors.
\label{tableResults}
\end{table}

Second, the number of iterations required to reach convergence strongly depends on the involved nonlinear laws. For the $h$- and $h$-$\phi$-formulations, the large number of iterations is due to the fixed point iterations on the permeability of the FM. We observed that in some cases, a Newton-Raphson scheme (with or without relaxation factors) applied on the permeability works without difficulty with a CPU time similar to that of coupled formulations, but this is not guaranteed in general. By contrast, for the $\bar a$- and $a$-formulations, the conductivity in HTS is significantly more difficult to handle. We only obtained convergence with a fixed point method.

For the coupled formulations, in surface and in volume, the number of iterations is directly reduced thanks to the use of the Newton-Raphson method on $\rho$ and $\nu$, without needing any parameter tuning.

Surface-coupled formulations appear to be the most efficient choices, especially the $h$-$\phi$-$a$. Volume-coupled formulations introduce more DOFs but may possibly be simpler to implement. Note that for modeling homogenized HTS-FM hybrids, e.g., stack of tapes, the volume coupling approach could be a convenient choice.

Interestingly, the CPU time per iteration does not scale directly with the number of DOFs. The matrix structures associated with the formulations are different and this also influences the linear solver resolution. Further investigations would provide a better understanding of this numerical behavior.





\section{Conclusion}

In this work, we compared the relevance of several finite element formulations for modeling 3D systems with high-temperature superconductors and ferromagnetic materials. To deal with the associated nonlinearities, the most efficient choice in terms of CPU time was a coupled \hpaf with surface coupling. While ensuring accurate results, this formulation combines a good robustness and a low number of degrees of freedom, thus leading to efficient simulations.


\section*{Acknowledgment}

Computational resources are provided by the Consortium des Équipements de Calcul Intensif (CÉCI), funded by the Fonds de la Recherche Scientifique de Belgique (F.R.S.-FNRS) under Grant No. 2.5020.11. J. Dular is a research fellow funded by the F.R.S-FNRS.

\bibliographystyle{ieeetr}
\bibliography{paperReferences}

\end{document}